\documentclass[aps,prl,twocolumn,superscriptaddress,showpacs,preprintnumbers,amsmath,amssymb]{revtex4}

\usepackage{orcidlink} 

\usepackage{afterpage,rotating}
\usepackage{currvita}
\usepackage{tabularx}
\usepackage{nonfloat}
\usepackage{graphicx}
\usepackage{dcolumn}
\usepackage{bm}
\usepackage{subfigure}
\usepackage{mathrsfs}
\usepackage{amsmath,amsbsy,amssymb,lmodern}
\usepackage{epsfig,color}
\usepackage{placeins}
\usepackage{rccol}
\usepackage{comment}
\usepackage{hyperref} 
\usepackage{setspace} 
\usepackage{tabularx} 
\usepackage{multirow}
\usepackage{braket}
\usepackage{accents}
\usepackage{xcolor}

\graphicspath{{figs}}

\DeclareRobustCommand\mybar[1]{\accentset{\rule{0.5em}{0.6pt}}{#1}}

\begin{document}




\preprint{\vbox{ \hbox{ }
\hbox{Belle Preprint 2024-01 \hfill}
\hbox{KEK Preprint 2023-48 \hfill}
}}



\title{ \quad\\[1.0cm] Search for Baryon-Number-Violating Processes in $B^-$ Decays to the $\mybar{\Xi}_{c}^{0} \mybar{\Lambda}_{c}^{-}$ Final State}


\noaffiliation
  \author{T.~Gu\,\orcidlink{0000-0002-1470-6536}} 
  \author{V.~Savinov\,\orcidlink{0000-0002-9184-2830}} 
  \author{I.~Adachi\,\orcidlink{0000-0003-2287-0173}} 
  \author{H.~Aihara\,\orcidlink{0000-0002-1907-5964}} 
  \author{D.~M.~Asner\,\orcidlink{0000-0002-1586-5790}} 
  \author{H.~Atmacan\,\orcidlink{0000-0003-2435-501X}} 
  \author{T.~Aushev\,\orcidlink{0000-0002-6347-7055}} 
  \author{R.~Ayad\,\orcidlink{0000-0003-3466-9290}} 
  \author{Sw.~Banerjee\,\orcidlink{0000-0001-8852-2409}} 
  \author{K.~Belous\,\orcidlink{0000-0003-0014-2589}} 
% ...
\collaboration{The Belle Collaboration}



\begin{abstract}
\par We report the results of the first search 
for $B^-$ decays to the $\mybar{\Xi}_{c}^{0} \mybar{\Lambda}_{c}^{-}$ final state 
using 711~${\rm fb^{-1}}$ of data collected at the $\Upsilon(4S)$ resonance
with the Belle detector at the KEKB asymmetric-energy $e^+ e^-$ collider. 
The results are interpreted in terms of 
both 
direct baryon-number-violating $B^-$ decay 
and 
$\Xi_{c}^{0}-\mybar{\Xi}_{c}^{0}$ oscillations 
which follow the Standard Model decay 
$B^- \to \Xi_{c}^{0} \mybar{\Lambda}_{c}^{-}$. 
We observe no evidence for baryon number violation 
and set the 95\% confidence-level upper limits 
on the ratio of baryon-number-violating and 
Standard Model branching fractions 
$
{\mathcal{B}(B^- \rightarrow \mybar{\Xi}_{c}^{0} \mybar{\Lambda}_{c}^{-})}/
{\mathcal{B}(B^- \rightarrow \Xi_{c}^{0} \mybar{\Lambda}_{c}^{-})}$
to be 
$< 2.7\%$ 
and 
on 
the $\Xi_{c}^{0} - \mybar{\Xi}_{c}^{0}$ oscillation angular frequency 
$\omega$ 
to be $< 0.76\ \mathrm{ps}^{-1}$ (equivalent to $\tau_{\rm mix} > 1.3$~ps).
\end{abstract}



\maketitle

\tighten

{\renewcommand{\thefootnote}{\fnsymbol{footnote}}}
\setcounter{footnote}{0}



Understanding the origin of 
the matter-antimatter asymmetry of the universe 
is 
one of the greatest challenges in particle physics. 
Three conditions necessary for baryogenesis, 
a hypothesized physical process necessary 
for generating such asymmetry in the early universe 
are (1) baryon number violation (BNV), 
(2) $C$ and $CP$ violation, and 
(3) departure from thermal/chemical equilibrium~\cite{Sakharov:Baryogenesis}. 
Of these three Sakharov conditions, 
experimental evidence for $CP$ violation has been observed  
and departure from thermal equilibrium could, 
in certain extensions of the Standard Model (SM), 
be satisfied through the expansion of the universe. 
%
%
No experimental evidence for BNV has been obtained so far. 


A variety of processes can be used to search for BNV, 
{\it e.g.}, 
proton decay~\cite{ProtonDecay}, 
which was originally proposed as a way to probe physics 
at the energy scale of grand unification, 
and direct BNV decays 
of the $\tau$ lepton~\cite{CLEO:tauDecay,Belle:tauDecay} 
and $B$ mesons~\cite{BABAR:Bmeson}, 
whose rates are usually extremely suppressed~\cite{Hou:2005iu}. 
Most of such BNV processes would be mediated by transitions 
which violate two discrete quantum numbers, 
baryon number $B$ and lepton number $L$, 
but conserve the difference 
$\Delta(B-L)$ between them. 
Both $B$ and $L$ numbers are, 
from the perspective of the SM, 
accidental, i.e., not protected by gauge symmetries, 
and could be violated non-perturbatively 
at high temperatures in the early universe~\cite{Kuzmin:1985mm}. 
Existing experimental data strongly constrain 
new physics in such $\Delta(B-L)=0$ processes. 
The exploration of the BNV landscape has recently been expanded 
to the domain of $\Delta(B-L) = 2$ processes via 
baryon-antibaryon oscillations. 
The flagship effort, 
motivated by the discovery of neutrino oscillations 
which require $\Delta(B-L) = 2$ interactions in the seesaw mechanism~\cite{min,moh,tao}, 
is in the area of neutron-antineutron oscillations~\cite{NeutronOsci}. 
Recently, the BES~III experiment 
extended the search for $\Delta(B-L) = 2$ processes 
to include the $s$ quark 
via $\Lambda^0 - \bar{\Lambda}^0$ oscillations 
and obtained a stringent limit~\cite{BESIII:2023tge}  
on the time constant of such oscillations to be 
$\tau_{\rm mix} > 1.7 \times 10^{-7}$~s 
at the 90\% confidence level (C.L.). 
The LHCb experiment performed a similar search~\cite{LHCb:Xib0Osci} 
for $\Xi_{b}^{0} - \mybar{\Xi}_{b}^{0}$ oscillations 
in the bottom sector 
and set the 95\%~C.L. upper limit on the oscillation rate to be 
$\omega<0.08~\rm{ps^{-1}}$ (equivalent to $\tau_{\rm mix} = 1/\omega > 13$~ps).  
In this Letter we expand the study of baryon-antibaryon oscillations 
to include the charm sector and report the results of the first search 
for BNV processes in $B^-$ decays 
to the $\mybar{\Xi}_{c}^{0} \mybar{\Lambda}_{c}^{-}$ final state. 
The unique feature of the analysis presented here 
is our ability to probe $\Delta(B-L) = 2$ processes 
which can proceed through several pathways. 
Such BNV transitions could be due to the direct BNV decay of $B^-$, 
which, similarly to BNV in $\tau$ lepton decays, is likely to be extremely suppressed.  
Also, they could be the result of the SM decay of $B^-$ 
followed by two possible scenarios associated with $\Xi_{c}^{0}$: 
the direct BNV process 
and 
$\Xi_{c}^{0}-\mybar{\Xi}_{c}^{0}$ oscillations, 
which were recently proposed~\cite{Theory:CharmBaryonOsci} 
as a possible new mechanism for baryogenesis. 
%
%
%
%
%

The connection between the two main BNV hypotheses considered in this Letter can be further clarified as follows. 
Since charmed baryons have a relatively short lifetime (e.g., the $\Xi_{c}^{0}$ lifetime is $0.152$~ps~\cite{PDG2022}), 
we are not able to resolve their decay vertices with the Belle detector~\cite{BelleDetector,Belle:2012iwr}. 
Therefore, from the analysis perspective, 
the SM decay $B^- \rightarrow \Xi_{c}^{0} \mybar{\Lambda}_{c}^{-}$
followed by the oscillation of $\Xi_{c}^{0}$ into $\mybar{\Xi}_{c}^{0}$ 
(or direct BNV decay of $\Xi_{c}^{0}$) 
is indistinguishable from the direct BNV decay $B^- \rightarrow \mybar{\Xi}_{c}^{0} \mybar{\Lambda}_{c}^{-}$. 

We measure 
the ratio between $B^-$ decay rates 
for the 
$\Xi_{c}^{0} \mybar{\Lambda}_{c}^{-}$ 
and 
$\mybar{\Xi}_{c}^{0} \mybar{\Lambda}_{c}^{-}$ 
final states 
and interpret this result 
as the ratio between 
branching fractions 
for direct BNV and SM decays. 
To address the charmed baryon-antibaryon oscillation hypothesis, 
assuming that the BNV decay of $B^-$ is actually 
the previously observed~\cite{Belle:Xic0BR} SM decay 
$B^- \to \Xi_{c}^{0} \mybar{\Lambda}_{c}^{-}$ 
followed by the non-SM 
$\Xi_{c}^{0} - \mybar{\Xi}_{c}^{0}$ oscillations, 
we measure their frequency.
In our study, 
the final states 
$\Xi_{c}^{0} \mybar{\Lambda}_{c}^{-}$ and $\mybar{\Xi}_{c}^{0} \mybar{\Lambda}_{c}^{-}$ 
%
%
are referred to as the SM and BNV modes, respectively.
Charge conjugate modes are included throughout this Letter. 


%
%

The $\Xi_{c}^{0}$ and $\mybar{\Xi}_{c}^{0}$ baryons are produced as flavor eigenstates, 
and then 
evolve and decay as superpositions of eigenstates of the Hamiltonian. 
The time evolution depends on the mixing parameters $x = (M_1 - M_2) / \Gamma$ and $y = (\Gamma_1 - \Gamma_2) / 2 \Gamma$, 
where $M_{1,2}$ and $\Gamma_{1,2}$ are the masses and widths of the eigenstates and $\Gamma = (\Gamma_1 + \Gamma_2)/2$.
Assuming no $CP$ violation and small mixing parameters, 
the time evolution of the event rate ratio between BNV and SM decays of a $\Xi_{c}^{0}$ state 
is described by the standard mixing formalism~\cite{Lenz:2020awd} as 
\begin{equation}
r(t) = 
\left( R_D + \sqrt{R_D}y'\Gamma t + \frac{x'^2+y'^2}{4}\Gamma^2 t^2 \right) e^{-\Gamma t},
\label{eq:formula_1}
\end{equation}

\noindent where $R_D$ is the ratio between branching fractions of $\Xi_{c}^{0}$ for direct BNV and SM modes, 
$x' = x \cos\delta + y \sin\delta$, $y' = -x \sin\delta + y \cos\delta$,
and $\delta$ is the strong phase difference between direct BNV and SM decays (with mixing). 
The time-integrated ratio between decay rates for the BNV and SM modes is described by 
\begin{equation}
R = R_D + \sqrt{R_D}y' + \frac{x'^2+y'^2}{2}.
\label{eq:formula_2}
\end{equation}

Assuming the $\Xi_{c}^{0} - \mybar{\Xi}_{c}^{0}$ oscillation hypothesis only and 
no direct BNV decay of $\Xi_{c}^{0}$ (i.e., $R_D = 0$), 
the time-integrated ratio of the decay rates for the BNV and SM modes is given by 
\begin{equation}
R = 2 \left[\left(\frac{\Delta M}{2}\right)^2 + \left(\frac{\Delta \Gamma}{4}\right)^2 \right] \tau^2 = 2 \omega^2 \tau^2, 
\label{eq:formula_3}
\end{equation}
\noindent where $\Delta M = M_1 - M_2$, $\Delta \Gamma = \Gamma_1 - \Gamma_2$, $\omega$ is the angular frequency of $\Xi_{c}^{0} - \mybar{\Xi}_{c}^{0}$ oscillations and $\tau$ is the lifetime of $\Xi_{c}^{0}$. 


This analysis is based on the full data sample of 711~${\rm fb^{-1}}$ 
collected at the $\Upsilon(4S)$ resonance with the Belle detector 
at the KEKB asymmetric-energy $e^+ e^-$ collider~\cite{KEKB}.
The Belle detector is a large-solid-angle magnetic
spectrometer that consists of a silicon vertex detector (SVD),
a 50-layer central drift chamber (CDC), an array of
aerogel threshold Cherenkov counters (ACC),
a barrel-like arrangement of time-of-flight
scintillation counters (TOF), and an electromagnetic calorimeter
comprised of CsI(Tl) crystals (ECL) located inside 
a super-conducting solenoid coil that provides the aforementioned 1.5~T magnetic field. 
The effect 
of the magnetic field 
on the energy splitting 
of the baryon and anti-baryon states 
can be safely ignored. 
An iron flux-return located outside of
the coil is instrumented to detect $K_L^0$ mesons and to identify
muons (KLM).  The detector is described in detail elsewhere~\cite{BelleDetector,Belle:2012iwr}.

Several signal Monte Carlo (MC) samples are generated 
to develop the signal selection criteria 
and estimate the reconstruction efficiencies. 
The MC generators 
{\sc EvtGen}~\cite{EvtGen}, 
{\sc PHOTOS}~\cite{Barberio:1993qi} 
and
{\sc PYTHIA}~\cite{Sjostrand:2007gs}
are used to simulate 
hadronic decay processes, 
final state radiation 
and hadronization, 
respectively. 
The {\sc GEANT3}~\cite{GEANT3} toolkit 
is used to model the detector response. 
To study backgrounds 
we use an MC sample of $\Upsilon(4S) \rightarrow B \overline{B}$ 
and $e^{+}e^{-} \rightarrow q \bar{q}$ hadronic continuum events 
with $q=u,d,s,c$ at $\sqrt{s}=10.58$~GeV 
corresponding to six times the integrated luminosity of the Belle data. 
Both background sources contribute to the events remaining after applying all selection criteria. 

In our analysis, 
the $\Xi_{c}^{0}$ is reconstructed in three decay channels ($\Xi^{-} \pi^{+}$, $\Lambda^{0} K^{-} \pi^{+}$ and $p K^{-} K^{-} \pi^{+}$), 
and the $\mybar{\Lambda}_{c}^{-}$ is reconstructed in two decay channels ($\overline{p} K_{S}^{0}$ and $\overline{p} K^{+} \pi^{-}$). 
Thus a total of six decay channels of $B^-$ mesons are analyzed.
The $\Xi^{-}$, $\Lambda^0$ and $K_S^0$ candidates 
are reconstructed via 
$\Xi^{-} \rightarrow \Lambda^0 \pi^{-}$, 
$\Lambda^0 \rightarrow p \pi^{-}$ 
and 
$K_S^0 \rightarrow \pi^+\pi^-$ 
decays, 
respectively. 

Final state charged particles are required to have transverse momenta 
(in the plane perpendicular to the direction of the $e^+$ beam) 
above 50~MeV/$c$. 
To identify them 
we use information from the CDC, TOF, and ACC 
and prepare particle identification (PID) likelihoods~\cite{BellePID} 
$L_i$ for particle species $i = K, \pi, p$. 
Distinct likelihoods that also include ECL information 
are used to distinguish electron ($e$) and non-electron ($h$) hypotheses. 
We use the ratios of the PID likelihoods $R_{i/j} = L_{i}/(L_{i}+L_{j})$ 
to select signal event candidates. 
%
%
Pions, kaons and protons are identified 
by requiring 
$R_{\pi/K}>0.6$ and $R_{e/h}<0.95$, 
$R_{\pi/K}<0.6$ and $R_{e/h}<0.95$, 
and 
$R_{p/K}>0.6$ and $R_{p/\pi}>0.6$, 
respectively. 
No such requirements are applied to 
the particles from $K_{S}^{0}$ and $\Lambda^0$ decays.
The PID efficiency 
depends on the particle species and kinematics 
and varies between 92\% and 98\%. 
PID misidentification rate 
for hadrons 
is between 4\% and 6\% per particle. 
%
%

$K_{S}^{0}$ and $\Lambda^0$ candidates are reconstructed 
in a multivariate analysis 
using a neural network technique~\cite{NeuroBayes,NakanoNeuroBayes}, 
and 
a kinematic fit to their decay vertices is performed to improve the mass resolution. 
The reconstructed masses of the $\Lambda^0$ and $K_S^0$ candidates are required 
to be within 10~MeV/$c^2$ ($\approx \pm 5 \sigma$) and 30~MeV/$c^2$ ($\approx \pm 10 \sigma$) 
of the nominal $\Lambda^0$ and $K_S^0$ masses~\cite{PDG2022}. 
These requirements reject around 20\% of background events and are 
at least 98\%-efficient 
for signal events. 

For each of the intermediate particle candidates 
($K_S^0, \Lambda^0, \Xi^{-}, \Xi_{c}^{0}, \mybar{\Lambda}_{c}^{-}$), 
the tracks reconstructed for its daughter particles are refit to a common vertex and 
their invariant mass is constrained to the nominal value. 
The momenta and decay vertices obtained from such constraints 
are then used in the parent particle reconstruction. 
The fit quality $\chi^2/n.d.f.$ is required to be smaller than 100 
for each individual fit, where $n.d.f.$ is the number of degrees of freedom. 
This mass-vertex $\chi^2$-based requirement suppresses the background by a factor of 3.
We apply the invariant mass requirements 
$|M_{\Xi_{c}^{0}} - m_{\Xi_{c}^{0}}| < 20~{\rm MeV}/c^2$, 
$|M_{\mybar{\Lambda}_{c}^{-}} - m_{\mybar{\Lambda}_{c}^{-}}| < 10~{\rm MeV}/c^2$ and 
$|M_{\Xi^{-}} - m_{\Xi^{-}}| < 10~{\rm MeV}/c^2$ ($\approx 3 \sigma$ for each), 
where $M_{\Xi_{c}^{0}}$, $M_{\mybar{\Lambda}_{c}^{-}}$ and $M_{\Xi^{-}}$ 
are the reconstructed masses of 
$\Xi_{c}^{0}$, $\mybar{\Lambda}_{c}^{-}$ and $\Xi^{-}$ candidates, 
and 
$m_{\Xi_{c}^{0}}$, $m_{\mybar{\Lambda}_{c}^{-}}$ and $m_{\Xi^{-}}$ are their nomimal masses~\cite{PDG2022}, respectively.
The requirements applied to the reconstructed invariant masses after kinematic fitting remove 90\% of the remaining background.

$B^-$ candidates are identified using 
the beam-energy constrained mass 
%
$M_{\rm bc} = \sqrt{(E_{\rm beam})^{2}-|\vec{p}_B|^2}$
and the energy difference $\Delta E = E_B-E_{\rm beam}$, where $E_{\rm beam}$ is the beam energy, 
%
$\vec{p}_B$ 
and $E_B$ are the reconstructed momentum and energy of the $B^-$ candidate, 
calculated in the $e^{+}e^{-}$ center-of-mass frame.
We require $M_{\rm bc}>$~5.20~GeV/$c^2$ and $|\Delta E| <$~0.25~GeV; 
the efficiency of this selection exceeds 99\%.
The signal region is defined as $M_{\rm bc}>$~5.27~GeV/$c^2$ and $|\Delta E| <$~0.02~GeV. 
The region $M_{\rm bc}>$~5.26~GeV/$c^2$ in the BNV analysis of data is blinded 
until the final fit to extract the branching fraction for the BNV mode is performed. 
The region $M_{\rm bc} \le 5.26$~GeV/$c^2$ defines the sideband. 

After applying all selection criteria, 
the percentages of reconstructed signal MC events that contain more than one candidate are, 
depending on the decay channel, 
between 6\% and 17\%. 
The candidate with the smallest cumulative $\chi^2$ obtained from 
the kinematic fits to $\Xi_{c}^{0}$, $\mybar{\Lambda}_{c}^{-}$ and $\Xi^{-}$ (when present in the decay chain) is selected as the best candidate.
Depending on the channel, the best candidate is correctly reconstructed 
in between 72\% and 94\% of signal MC events with more than one candidate. 
The signal candidate is correctly reconstructed 
in between 85\% and 97\% of events with at least one candidate.
Overall reconstruction efficiencies for individual channels are in the range 
between 6.6\% and 9.9\%. 
%

\begin{figure*}[!htb]
\hfill
    \centering

    \begin{minipage}[c]{0.49\linewidth}
    \centering
    \includegraphics[width=\textwidth]{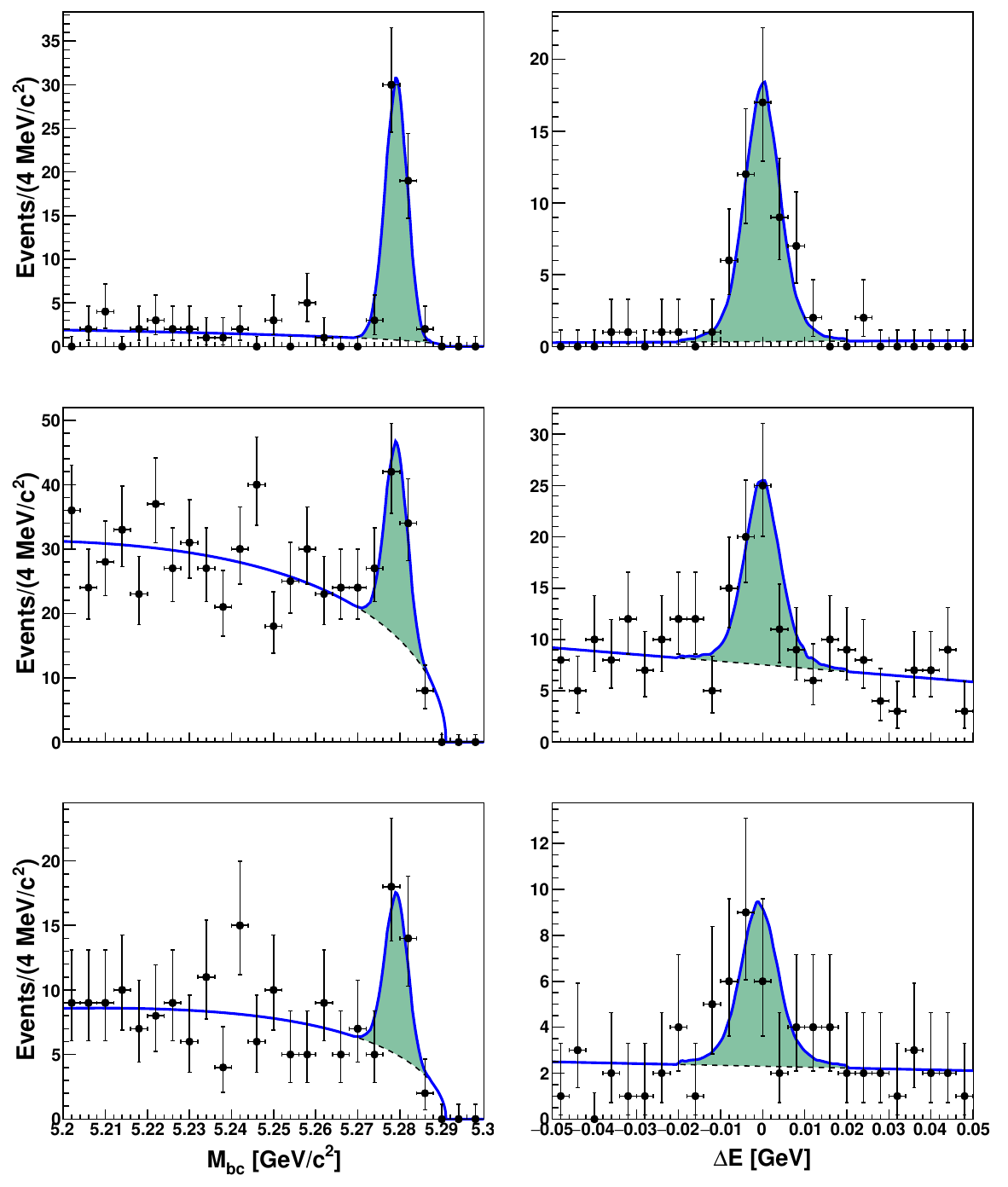}
    \end{minipage}
\hfill
\hspace{0mm}
\hfill
    \begin{minipage}[c]{0.49\linewidth}
    \centering
    \includegraphics[width=\textwidth]{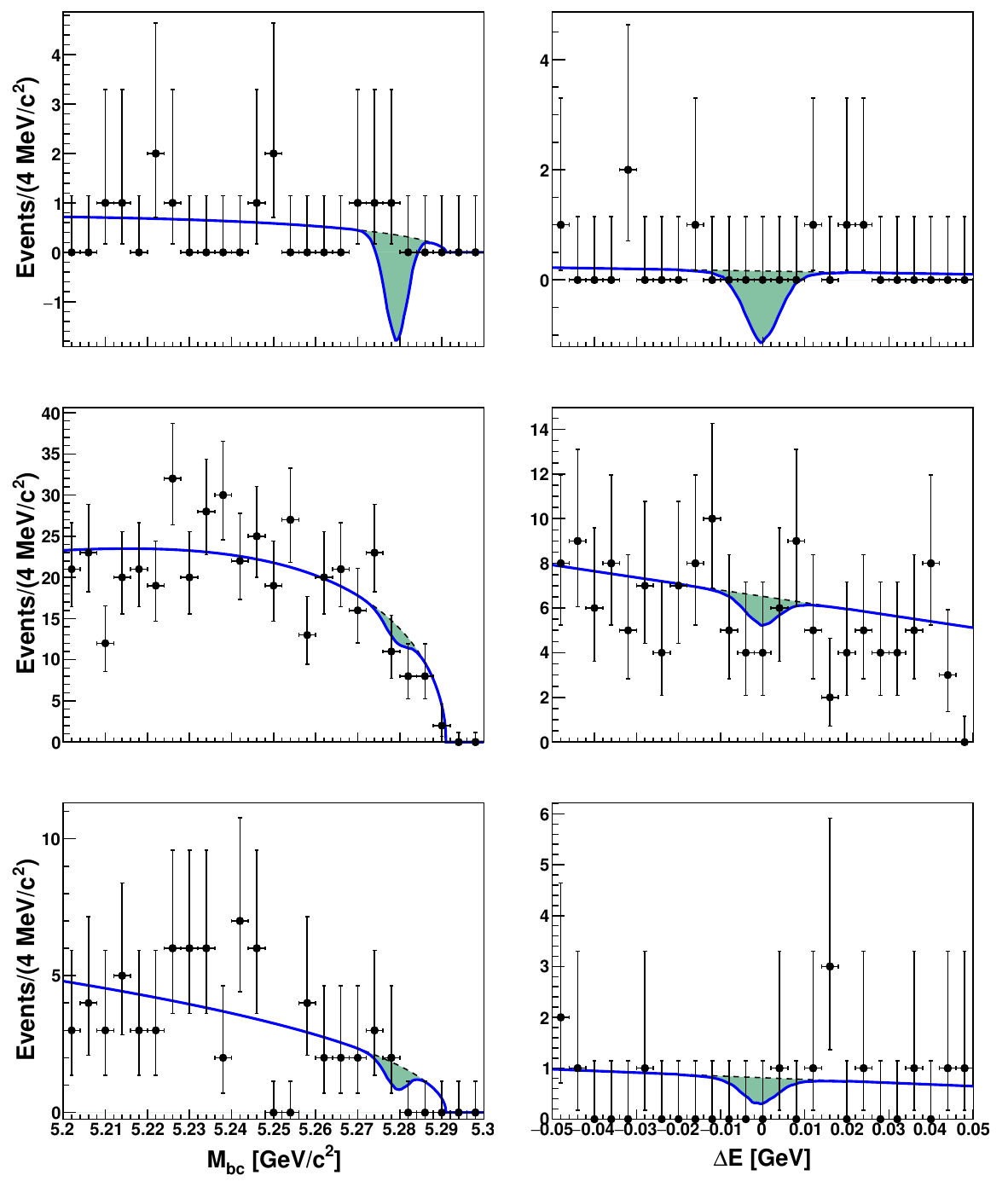}
    \end{minipage}
\hfill
    \caption{
    Signal-region projections of the data fit result 
    onto $M_{\rm bc}$ and $\Delta E$ for  
    (left) the SM mode 
    $B^- \rightarrow \Xi_{c}^{0} \mybar{\Lambda}_{c}^{-}$ 
    and
    (right) the BNV mode 
    $B^- \rightarrow \mybar{\Xi}_{c}^{0} \mybar{\Lambda}_{c}^{-}$ 
    in the 
    $\Xi_{c}^{0} \rightarrow \Xi^{-} \pi^{+}$, 
    $\Lambda^0 K^{-} \pi^{+}$ 
    and 
    $p K^{-} K^{-} \pi^{+}$ (top, middle and bottom) 
    channels, 
    summed over the two reconstructed $\mybar{\Lambda}_{c}^{-}$ decay modes. 
    Dots with error bars represent the binned data, 
    blue solid curves show the results of the fit, 
    green-filled regions and black dashed curves show the signal and background fit components, respectively.
}
    \label{fig:open_box_both}
\end{figure*}

We measure 
branching fractions 
$\mathcal{B}(B^- \rightarrow \Xi_{c}^{0} \mybar{\Lambda}_{c}^{-})$ and 
$\mathcal{B}(B^- \rightarrow \mybar{\Xi}_{c}^{0} \mybar{\Lambda}_{c}^{-})$ 
for the SM and BNV modes. 
%
For each of the two measurements, 
a 2D unbinned extended maximum likelihood fit is performed simultaneously to 
$M_{\rm bc}$ vs $\Delta E$ distributions for the SM (BNV) decay 
$B^- \rightarrow \Xi_{c}^{0} (\mybar{\Xi}_{c}^{0}) \mybar{\Lambda}_{c}^{-}$ 
in the $\Xi_{c}^{0} \rightarrow \Xi^{-} \pi^{+}$, $\Lambda^0 K^{-} \pi^{+}$ and $p K^{-} K^{-} \pi^{+}$ channels, 
summed over the two reconstructed $\mybar{\Lambda}_{c}^{-}$ decay modes. 
Branching fractions of $\Xi_{c}^{0}$ and reconstruction efficiencies for individual channels are used to fix the relative yields in the fit.
%
To handle signal correlations between $M_{\rm bc}$ and $\Delta E$, 
a 2D smoothed histogram~\cite{RooFit} 
obtained from signal MC samples is used to model the signal probability density function (PDF). 
Bin widths used for these histograms are 2~${\rm MeV}/c^2$ and 2.5~${\rm MeV}$ 
for $M_{\rm bc}$ and $\Delta E$, respectively. 
We use the second-order interpolation between the bins. 
The same signal PDFs are used for the SM and BNV modes. 
The 2D background PDF is assumed to be factorizable, 
i.e., $\mathcal{P}_{bkg}(M_{\rm bc}, \Delta E) =  \mathcal{P}_{bkg}(M_{\rm bc}) \times \mathcal{P}_{bkg}(\Delta E)$ 
as the correlations between $M_{\rm bc}$ and $\Delta E$ for background are found to be negligible. 
The background $M_{\rm bc}$ distribution is modeled with an ARGUS function~\cite{ARGUS} and 
the background $\Delta E$ distribution is modeled with a first-order Chebychev polynomial. 
No peaking backgrounds have been identified using MC samples. 
Background PDF parameters are not constrained in the fits. 
The fit results for the branching fractions for SM and BNV modes are 
    $\mathcal{B}(B^- \rightarrow \Xi_{c}^{0} \mybar{\Lambda}_{c}^{-}) = (1.13 \pm 0.12) \times 10^{-3}$ 
%
and
%
    $\mathcal{B}(B^- \rightarrow \mybar{\Xi}_{c}^{0} \mybar{\Lambda}_{c}^{-}) = (-7.78 \pm 2.70) \times 10^{-5}$, 
where only the statistical uncertainty is shown. 
According to toy MC experiments, the probability to obtain such or even more negative a result 
for the BNV mode is 25\%. 
Fig.~\ref{fig:open_box_both} shows the signal-region projections of the fit results to data onto 
the $M_{\rm bc}$ and $\Delta E$ distributions for SM and BNV analyses. 
The result for the SM mode is consistent with the previous measurement from Belle~\cite{Belle:Xic0BR}.
The fit results correspond to 
$46.6 \pm 4.9$ ($-3.2 \pm 1.1$), 
$49.6 \pm 5.3$ ($-3.4 \pm 1.2$) 
and 
$20.9 \pm 2.2$ ($-1.5 \pm 0.5$) 
events in the SM (BNV) modes 
$\Xi^{-} \pi^{+}$ ($\Xi^{+} \pi^{-}$), 
$\Lambda^0 K^{-} \pi^{+}$ ($\mybar{\Lambda}^0 K^{+} \pi^{-}$) 
and 
$p K^{-} K^{-} \pi^{+}$ ($\mybar{p} K^{+} K^{+} \pi^{-}$), 
respectively, 
where only the statistical uncertainty is shown. 
%
%
%
%
%
%
%
%
%
%
%
%
%
%
%


Since we use the same analysis procedure for SM and BNV decays, 
most of the systematic uncertainties, such as contributions from luminosity, PID selection, 
track reconstruction, $K_{S}^{0}$ and $\Lambda^0$ reconstruction, 
cancel in the ratio between branching fractions for the BNV and SM modes. 
The only significant contribution to systematic uncertainty is due to the PDF parameterization  
which is taken into account in the upper limit estimation procedure which is described later. 
The systematic uncertainties 
due to finite MC statistics 
and 
imperfect knowledge of the daughter particle branching fractions 
are 
0.4\% 
and 
0.02\%, 
respectively. 
The effect of these uncertainties on the final result is negligible.

\begin{figure}[!htb]
    \centering
    \begin{minipage}[t]{1\linewidth}
    \centering
    \includegraphics[width=0.95\textwidth]{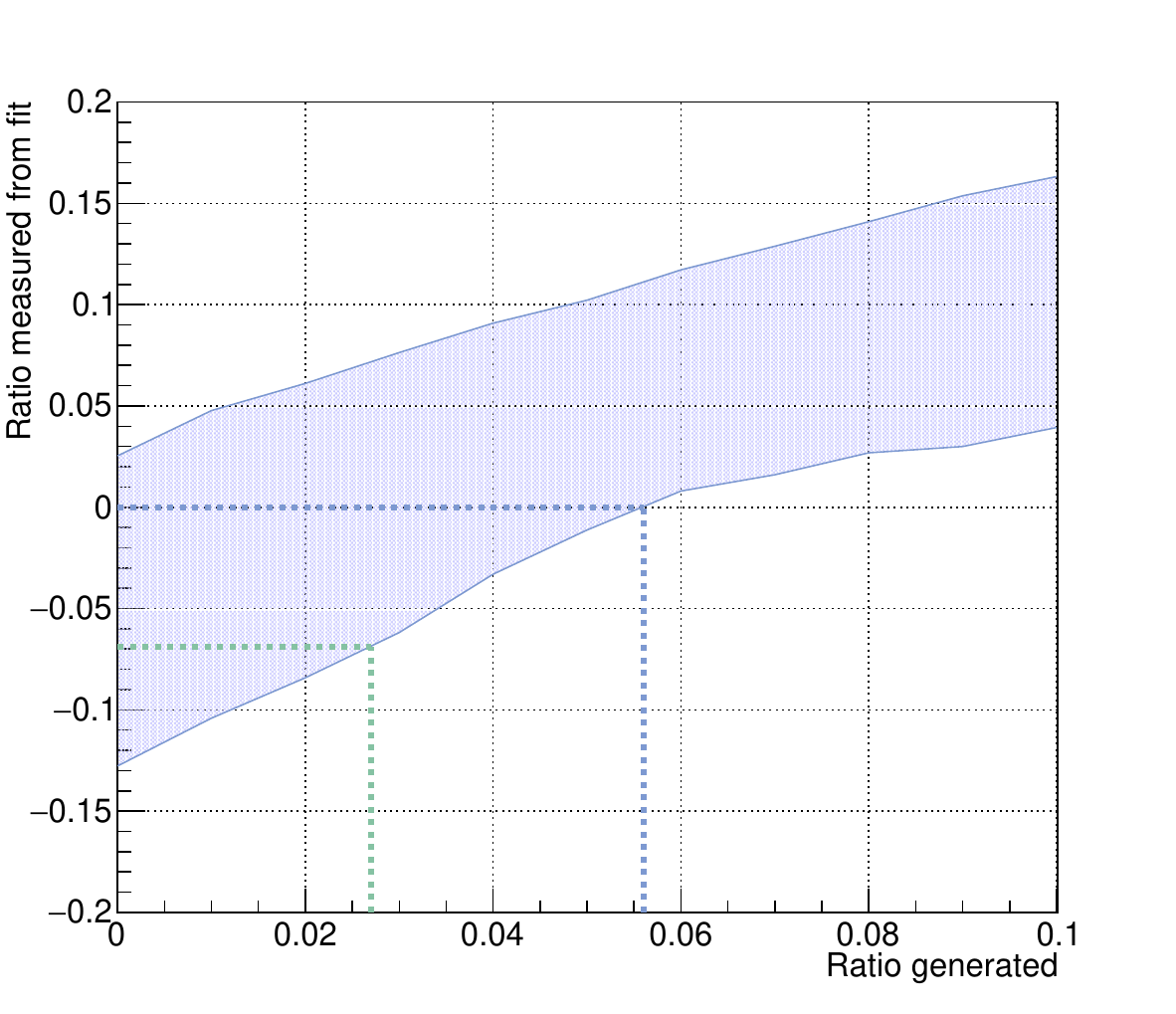}
    \end{minipage}
    \caption{The 90\%~C.L. belt for the ratio between branching fractions for the BNV and SM modes constructed including statistical and systematic uncertainties. Green and blue dotted lines demonstrate the procedures used to obtain our 95\%~C.L. upper limit and the sensitivity (using a zero result), respectively.}
    \label{fig:ratio_clband}
\end{figure}

To estimate the upper limit using the frequentist approach~\cite{NeymanCL}, 
we construct the 90\% C.L. belt for the ratio between the branching fractions for the BNV and SM modes. 
We perform 5000 pseudo-experiments for each assumed ratio 
and 
randomly sample the SM mode branching fraction based on its measured value and statistical uncertainty 
in each toy MC experiment. 
We use this procedure to estimate the number of signal events in each of the SM and BNV modes. 
The expected numbers of background events in SM and BNV modes 
are estimated using sideband data scaled using background MC.
Events are generated according to the fit models described previously.
Finally, to measure the ratio between branching fractions for the BNV and SM modes, 
we fit our model with PDFs, that are randomly varied 
to incorporate systematic uncertainties due to PDF parametrization.
To take into account the difference between data and MC resolution functions, 
the width of each signal PDF is modified using a scale factor 
randomly sampled from a Gaussian distribution with $\mu = 1$ and $\sigma = 0.1$, 
in order to increase or decrease the width of the signal PDFs by 10\%, on average. 
In order to include systematic uncertainties due to background PDF shapes, 
the background $M_{\rm bc}$ distribution is modeled with an ARGUS function with released threshold 
and 
the background $\Delta E$ distribution is modeled with a second-order Chebychev polynomial.
For each ensemble of pseudo-experiments, the lower and upper ends of respective confidence intervals correspond to 
the values for which 5\% of the fit results are below and above these values. 
Fig.~\ref{fig:ratio_clband} shows the 90\% C.L. belt for the ratio of branching fractions 
for the BNV and SM modes 
after including both statistical and systematic uncertainties. 

Based on the central value of $-0.069$ for the measured ratio 
between branching fractions for the SM and BNV modes, 
the upper limit on their ratio, 
$R = \mathcal{B}(B^- \rightarrow \mybar{\Xi}_{c}^{0} \mybar{\Lambda}_{c}^{-})/\mathcal{B}(B^- \rightarrow \Xi_{c}^{0} \mybar{\Lambda}_{c}^{-})$ 
is estimated to be  
$ < 2.7$\% at the 95\%~C.L. 

An alternative interpretation of our results is provided assuming that no direct BNV decay of $B^-$ takes place. 
In this case $R$ 
is the time-integrated ratio 
between $\Xi_{c}^{0}$ event rates 
for the BNV and SM modes 
given by Eq.~\ref{eq:formula_2}. 
Assuming no direct BNV in $\Xi_{c}^{0}$ decays allows us to use Eq.~\ref{eq:formula_3}
to estimate the upper limit on the oscillation angular frequency 
to be $ \omega < 0.76\ \mathrm{ps^{-1}}$ at the 95\%~C.L., equivalent to $\tau_{\rm mix} > 1.3$~ps. 

Assuming a zero result for the $B^-$ branching fraction for the BNV mode, 
the sensitivity for the ratio between branching fractions for the BNV and SM modes is $R=5.6\%$ at the 95\%~C.L. 
Under the hypothesis of $\Xi_{c}^{0} - \mybar{\Xi}_{c}^{0}$ oscillations 
a zero result corresponds to a sensitivity 
$ \omega = 1.10\ \mathrm{ps^{-1}}$ at the 95\%~C.L. 
for the oscillation angular frequency 
(equivalent to $\tau_{\rm mix} > 0.91$~ps). 

In summary, 
using the full data sample collected by the Belle experiment at the $\Upsilon(4S)$ resonance, 
we performed the first search for the baryon-number-violating processes in  
$B^-$ decays to the $\mybar{\Xi}_{c}^{0} \mybar{\Lambda}_{c}^{-}$ final state.
We observe no evidence for baryon number violation 
and set the 95\%~C.L. upper limit on the ratio 
between branching fractions for the BNV and SM modes in $B^-$ decays 
to be $<2.7\%$. 
Assuming no direct BNV transitions in $\Xi_{c}^{0}$ decays, 
we set the 95\%~C.L. upper limit  
on the $\Xi_{c}^{0} - \mybar{\Xi}_{c}^{0}$ oscillation angular frequency 
to be 
$< 0.76\ \mathrm{ps^{-1}}$  (equivalent to $\tau_{\rm mix} > 1.3$~ps). 
This is the first experimental result 
on oscillations in the charmed baryon sector. 
This work, based on data collected using the Belle detector, which was
operated until June 2010, was supported by 
the Ministry of Education, Culture, Sports, Science, and
Technology (MEXT) of Japan, the Japan Society for the 
Promotion of Science (JSPS), and the Tau-Lepton Physics 
Research Center of Nagoya University; 
the Australian Research Council including grants
DP210101900, 
DP210102831, 
DE220100462, 
LE210100098, 
LE230100085; 
Austrian Federal Ministry of Education, Science and Research (FWF) and
FWF Austrian Science Fund No.~P~31361-N36;
National Key R\&D Program of China under Contract No.~2022YFA1601903,
National Natural Science Foundation of China and research grants
No.~11575017,
No.~11761141009, 
No.~11705209, 
No.~11975076, 
No.~12135005, 
No.~12150004, 
No.~12161141008, 
and
No.~12175041, 
and Shandong Provincial Natural Science Foundation Project ZR2022JQ02;
the Czech Science Foundation Grant No. 22-18469S;
Horizon 2020 ERC Advanced Grant No.~884719 and ERC Starting Grant No.~947006 ``InterLeptons'' (European Union);
the Carl Zeiss Foundation, the Deutsche Forschungsgemeinschaft, the
Excellence Cluster Universe, and the VolkswagenStiftung;
the Department of Atomic Energy (Project Identification No. RTI 4002), the Department of Science and Technology of India,
and the UPES (India) SEED finding programs Nos. UPES/R\&D-SEED-INFRA/17052023/01 and UPES/R\&D-SOE/20062022/06; 
the Istituto Nazionale di Fisica Nucleare of Italy; 
National Research Foundation (NRF) of Korea Grant
Nos.~2016R1\-D1A1B\-02012900, 2018R1\-A2B\-3003643,
2018R1\-A6A1A\-06024970, RS\-2022\-00197659,
2019R1\-I1A3A\-01058933, 2021R1\-A6A1A\-03043957,
2021R1\-F1A\-1060423, 2021R1\-F1A\-1064008, 2022R1\-A2C\-1003993;
Radiation Science Research Institute, Foreign Large-size Research Facility Application Supporting project, the Global Science Experimental Data Hub Center of the Korea Institute of Science and Technology Information and KREONET/GLORIAD;
the Polish Ministry of Science and Higher Education and 
the National Science Center;
the Ministry of Science and Higher Education of the Russian Federation, Agreement 14.W03.31.0026, 
and the HSE University Basic Research Program, Moscow; 
University of Tabuk research grants
S-1440-0321, S-0256-1438, and S-0280-1439 (Saudi Arabia);
the Slovenian Research Agency Grant Nos. J1-9124 and P1-0135;
Ikerbasque, Basque Foundation for Science, and the State Agency for Research
of the Spanish Ministry of Science and Innovation through Grant No. PID2022-136510NB-C33 (Spain);
the Swiss National Science Foundation; 
the Ministry of Education and the National Science and Technology Council of Taiwan;
and the United States Department of Energy and the National Science Foundation.
These acknowledgements are not to be interpreted as an endorsement of any
statement made by any of our institutes, funding agencies, governments, or
their representatives.
We thank the KEKB group for the excellent operation of the
accelerator; the KEK cryogenics group for the efficient
operation of the solenoid; and the KEK computer group and the Pacific Northwest National
Laboratory (PNNL) Environmental Molecular Sciences Laboratory (EMSL)
computing group for strong computing support; and the National
Institute of Informatics, and Science Information NETwork 6 (SINET6) for
valuable network support.
We thank David McKeen (TRIUMF) and Brian Batell (University of Pittsburgh) for stimulating discussions. 
%



\begin{thebibliography}{10}

\bibitem{Sakharov:Baryogenesis}
A.~D.~Sakharov,
{Violation of CP invariance, C asymmetry, and baryon asymmetry of the universe,}
Sov. Phys. Usp. \textbf{34}, 392 (1991).

\bibitem{ProtonDecay}
R.~M.~Bionta \textit{et al.}, 
{The search for proton decay,}
AIP Conf. Proc. \textbf{123}, 321 (1984).

\bibitem{CLEO:tauDecay}
R.~Godang \textit{et al.} (CLEO Collaboration),
{Search for baryon and lepton number violating decays of the tau lepton,}
Phys. Rev. D \textbf{59}, 091303 (1999).


\bibitem{Belle:tauDecay}
D.~Sahoo \textit{et al.} (Belle Collaboration),
{Search for lepton-number- and baryon-number-violating tau decays at Belle,}
Phys. Rev. D \textbf{102}, 111101 (2020). 

\bibitem{BABAR:Bmeson}
P.~del Amo Sanchez \textit{et al.} (BABAR Collaboration),
{Searches for the baryon- and lepton-number violating decays $B^0\rightarrow\Lambda_c^+\ell^-$, $B^-\rightarrow\Lambda\ell^-$, and $B^-\rightarrow\bar{\Lambda}^0\ell^-$,}
Phys. Rev. D \textbf{83}, 091101 (2011).

\bibitem{Hou:2005iu}
W.~S.~Hou, M.~Nagashima and A.~Soddu,
{Baryon number violation involving higher generations,}
Phys. Rev. D \textbf{72}, 095001 (2005). 

\bibitem{Kuzmin:1985mm}
V.~A.~Kuzmin, V.~A.~Rubakov and M.~E.~Shaposhnikov, 
{On the Anomalous Electroweak Baryon Number Nonconservation in the Early Universe,}
Phys. Lett. B \textbf{155}, 36 (1985).

\bibitem{min}
P.~Minkowski,
{On the Spontaneous Origin of Newton's Constant,}
Phys. Lett. B \textbf{71}, 419 (1977). 


\bibitem{moh}
R.~N.~Mohapatra and G.~Senjanovic,
{Neutrino Masses and Mixings in Gauge Models with Spontaneous Parity Violation,}
Phys. Rev. D \textbf{23}, 165 (1981). 

\bibitem{tao}
T.~Han, H.~E.~Logan, B.~Mukhopadhyaya and R.~Srikanth,
{Neutrino masses and lepton-number violation in the littlest Higgs scenario,}
Phys. Rev. D \textbf{72}, 053007 (2005).


\bibitem{NeutronOsci}
D.~G.~Phillips II \textit{et al.}, 
{Neutron-antineutron oscillations: theoretical status and experimental prospects,}
Phys. Rept. \textbf{612}, 1 (2016).

\bibitem{BESIII:2023tge} 
M.~Ablikim \textit{et al.} (BES~III Collaboration), 
{Search for ${\bar\Lambda}-\Lambda$ baryon-number-violating oscillations in the decay $J/\psi \to pK^-\Lambda$ + $c.c.$,} 
Phys. Rev. Lett. \textbf{131}, 121801 (2023). 

\bibitem{LHCb:Xib0Osci}
R.~Aaij \textit{et al.} (LHCb Collaboration),
{Search for baryon-number violating ${\mathrm{\Xi}}_{b}^{0}$ oscillations,}
Phys. Rev. Lett. \textbf{119}, 181807 (2017).

\bibitem{Theory:CharmBaryonOsci}
K.~Aitken, D.~McKeen, T.~Neder and A.~E.~Nelson,
{Baryogenesis from oscillations of charmed or beautiful baryons,}
Phys. Rev. D \textbf{96}, 075009 (2017).



\bibitem{PDG2022}
R.~L.~Workman \textit{et al.} (Particle Data Group),
{Review of particle physics,}
PTEP \textbf{2022}, 083C01 (2022).

\bibitem{BelleDetector}
A.~Abashian \textit{et al.} (Belle Collaboration),
{The Belle detector,}
Nucl. Instrum. Meth. A \textbf{479}, 117 (2002).

\bibitem{Belle:2012iwr}
J.~Brodzicka \textit{et al.} (for the Belle Collaboration),
{Physics achievements from the Belle experiment,}
PTEP \textbf{2012}, 04D001 (2012).

\bibitem{Belle:Xic0BR}
Y.~B.~Li \textit{et al.} (Belle Collaboration),
{First measurements of absolute branching fractions of the $\Xi_c^0$ baryon at Belle,}
Phys. Rev. Lett. \textbf{122}, 082001 (2019).

\bibitem{Lenz:2020awd}
A.~Lenz and G.~Wilkinson, 
{Mixing and CP Violation in the Charm System,} 
Ann. Rev. Nucl. Part. Sci. \textbf{71}, 59 (2021).




\bibitem{KEKB}
S.~Kurokawa and E.~Kikutani,
{Overview of the KEKB accelerators,}
Nucl. Instrum. Meth. A \textbf{499}, 1 (2003),  
and other papers included in this Volume;
T.~Abe {\it et al.}, 
{Achievements of KEKB,}
PTEP \textbf{2013}, 03A001 (2013), 
and references therein.

\bibitem{EvtGen}
D.~J.~Lange,
{The {\sc EvtGen} particle decay simulation package,}
Nucl. Instrum. Meth. A \textbf{462}, 152 (2001).

\bibitem{Barberio:1993qi} 
E.~Barberio and Z.~Was,
{{\sc PHOTOS}: A Universal Monte Carlo for QED radiative corrections. Version 2.0,}
Comput. Phys. Commun. \textbf{79}, 291 (1994).

\bibitem{Sjostrand:2007gs}
T.~Sjostrand, S.~Mrenna and P.~Z.~Skands, 
{A Brief Introduction to {\sc PYTHIA} 8.1,} 
Comput. Phys. Commun. \textbf{178}, 852 (2008). 


\bibitem{GEANT3}
R.~Brun {\it et al.}, {GEANT Detector Description and Simulation Tool}, CERN ebook 10.17181/CERN.MUHF.DMJ1, (1994). 

\bibitem{BellePID}
E.~Nakano,
{Belle PID,}
Nucl. Instrum. Meth. A \textbf{494}, 402 (2002).

\bibitem{NeuroBayes}
M.~Feindt and U.~Kerzel,
{The NeuroBayes neural network package,}
Nucl. Instrum. Meth. A \textbf{559}, 190 (2006).


\bibitem{NakanoNeuroBayes}
H.~Nakano \textit{et al.} (Belle Collaboration),
{Measurement of time-dependent $CP$ asymmetries in $B^{0}\to K_S^0 \eta \gamma$ decays,}
Phys. Rev. D \textbf{97}, 092003 (2018).

\bibitem{RooFit}
W.~Verkerke and D.~P.~Kirkby,
{The RooFit toolkit for data modeling,} 
eConf \textbf{C0303241}, MOLT007 (2003). 

\bibitem{ARGUS}
H.~Albrecht \textit{et al.} (ARGUS Collaboration),
{Search for hadronic $b \to u$ decays,}
Phys. Lett. B \textbf{241}, 278 (1990).

\bibitem{NeymanCL}
J.~Neyman,
{Outline of a theory of statistical estimation based on the classical theory of probability,}
Phil. Trans. Roy. Soc. Lond. A \textbf{236}, 333 (1937).



\end{thebibliography}

\end{document}